\documentstyle[11pt,newpasp,twoside,psfig]{article}
\def\gs{\mathrel{\raise0.35ex\hbox{$\scriptstyle >$}\kern-0.6em
\lower0.40ex\hbox{{$\scriptstyle \sim$}}}}
\def\ls{\mathrel{\raise0.35ex\hbox{$\scriptstyle <$}\kern-0.6em
\lower0.40ex\hbox{{$\scriptstyle \sim$}}}}

\def\edcomment#1{\iffalse\marginpar{\raggedright\sl#1\/}\else\relax\fi}
\reversemarginpar

\begin{document}
\title{A Hubble Space Telescope Survey of X-ray Luminous 
Galaxy Clusters: Gravitationally Lensed Arcs and EROs}
\author{Graham P.\ Smith}
\affil{Department of Physics, University of Durham, South Road, Durham, 
DH1 3LE, UK}

\begin{abstract}
We are conducting a systematic lensing survey of X-ray luminous galaxy
clusters at  $z\sim0.2$ using the \emph{Hubble  Space Telescope (HST)}
and large ground-based telescopes.   We summarize initial results from
our survey,  including a  measurement of the  inner slope of  the mass
profile of  A\,383, and a search for  gravitationally lensed Extremely
Red Objects.
\end{abstract}

\section{Introduction}

Gravitational  lensing  by galaxy  clusters  is  a  powerful tool  for
studying the distribution of mass in galaxy clusters at $z>0.2$ (e.g.\
Kneib  et al.\  1996 --  K96; Smith  et al.\  2001a --  S01a)  and the
properties of high-redshift ($z\sim1$--6) galaxies (e.g. Smith et al.\
2002a --  S02a; Smail  et al.\  2001; Ellis et  al.\ 2001).   For this
reason,  significant   effort  was  invested  during   the  1990's  in
developing  the  lens   inversion  techniques  required  to  interpret
robustly  cluster  lensing  observations  (K96).  During  this  period
cluster lensing studies necessarily  concentrated on a small number of
well studied clusters.

Our survey builds on the pioneering work of the 1990's and applies the
K96  lens inversion  technique to  an objectively  selected  sample of
clusters.  Ideally we would select  our cluster sample based on direct
measurements of  their mass.  However,  in the absence of  large scale
weak lensing surveys, we rely on X-ray luminosity as a crude indicator
of cluster  mass for  the purpose of  sample selection.   We therefore
select    12   X-ray    luminous   clusters    ($L_{\rm   X}\ge8\times
10^{44}$\,erg\,s$^{-1}$, 0.1--2.4\,keV) in  a narrow redshift slice at
$z=0.17$--0.26, with  line of  sight reddening of  $E(B-V)\le0.1$ from
the XBACs sample (Ebeling et al.\ 1996).

We  describe  the  first  three  published results  from  our  survey:
detailed modeling  of the density  profile of A\,383 (S01a);  a search
for gravitationally  lensed Extremely  Red Objects (EROs)  (S02a); and
near-infrared (NIR)  spectroscopy of a  dusty ERO uncovered  in S02a's
survey (Smith et al.\ 2001b -- S01b).

\section{The Distribution of Matter in the Core ($r\ls20$kpc) of Abell 
383}

A\,383 was one  of the first clusters to be  observed by \emph{HST} as
part  of  our   survey,  revealing  numerous  previously  unidentified
multiple images  including a ``giant''  tangential arc and  two radial
arcs  (Fig.~1 \&  S01a).   Such  radial arcs  are  extremely rare  and
provide  a unique  opportunity to  measure  the slope  of the  cluster
density profile,  in contrast to  tangential arcs which are  much more
common and simply constrain the mass enclosed by the arc.

\smallskip
\noindent\begin{center}
\begin{minipage}{50mm}
\psfig{file=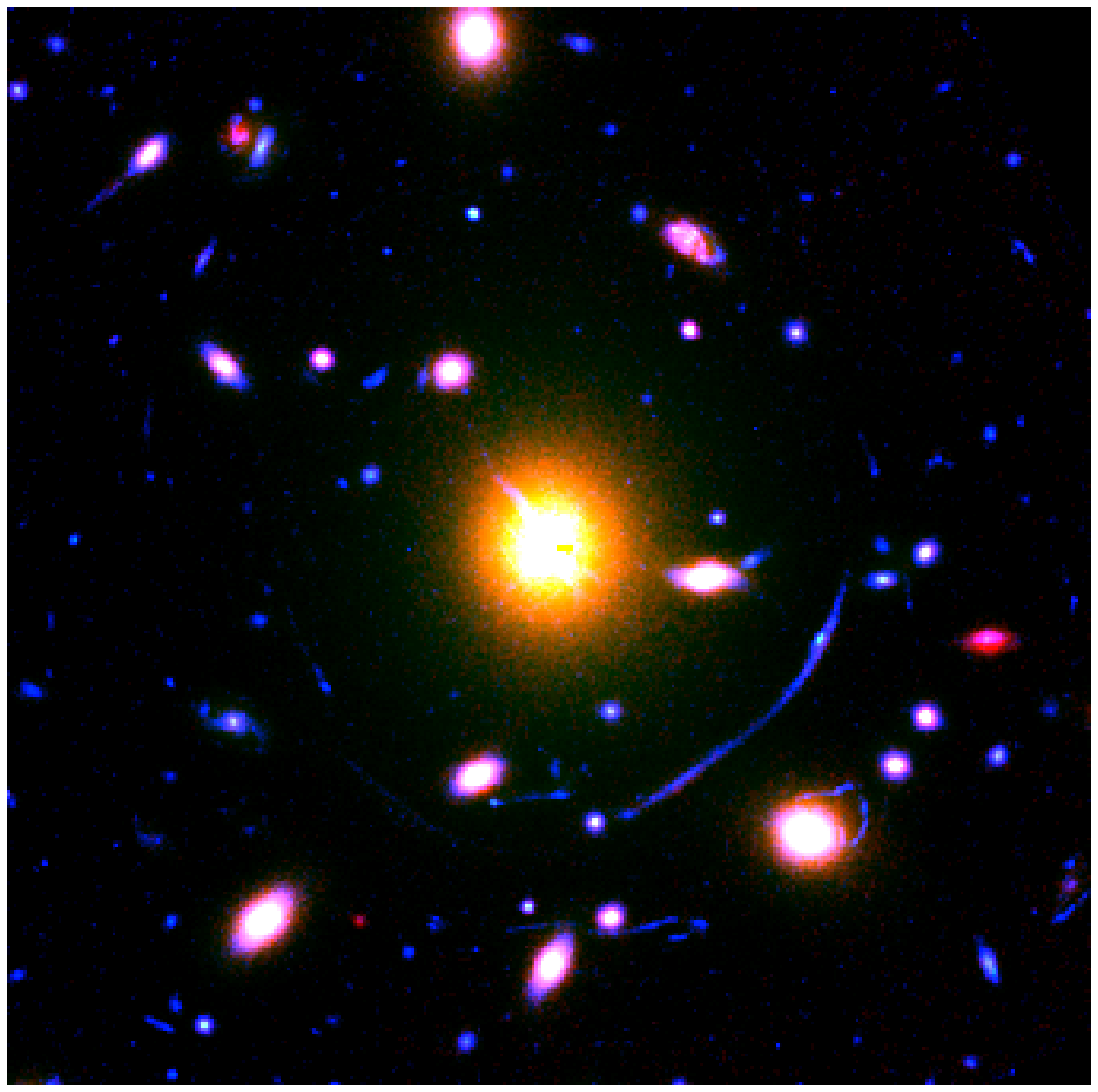,width=50mm,angle=0}
\end{minipage}
\hspace{10mm}
\begin{minipage}{50mm}
\psfig{file=fig1b.ps,height=50mm,angle=-90}
\end{minipage}
\end{center}
\smallskip
{\footnotesize{\noindent{\sc   Figure~1.}   --   Left:   The   central
$\sim50''\times50''$ of  A\,383, combining our  WFPC2/F702W frame with
$K$--band data from  UKIRT (see \S3).  The large  blue arc below-right
of the image-centre is a ``giant arc'' consisting of several images of
the same background  galaxy, including one spectroscopically confirmed
to lie at $z=1.01$.  The radial arcs (see text) lie in the very centre
of the cluster and are hidden  by the cluster cD galaxy in this frame.
Right: The radial  arcs in A\,383, as seen with  {\it HST} enhanced by
subtracting  a  median-smoothed frame  from  the  science frame.   The
bright emission at the bottom of this frame is the residual light from
the cD  galaxy after  subtracting off the  smoothed frame.   Each tick
mark represents $1''$.

}}
\medskip

Keck spectroscopy of the brightest feature of the giant arc reveals it
to be  an image  of a star  forming galaxy  at $z=1.01$.  We  use this
spectroscopic redshift  to calibrate a model of  the mass distribution
in this cluster and then  investigate the slope of the cluster density
profile  using  the radial  arcs.   Specifically,  we investigate  the
degeneracy  between the  various  model parameters  that describe  the
cluster-scale dark matter halo and the central galaxy in this cluster.
We  conclude that  a  central  galaxy mass  component  is required  to
explain  the   lensing  properties  of   the  cluster  and   that  the
de-projected slope of the inner  region of the density profile ($10\ls
r\ls50$kpc)  is $\alpha=-1.30\pm0.04$ where  $\rho\propto r^{\alpha}$.
The central slope of A\,383's  density profile therefore appears to be
intermediate between the predictions of Navarro, Frenk \& White (1997)
($\alpha=-1$) and the higher  resolution simulations of Ghigna et al.\
(1999) ($\alpha=-1.6\pm0.1$) (S01a).

Future analysis of A\,383  will fully map the systematic uncertainties
in  our lens  model,  including  those arising  from  the adoption  of
different parametrized  mass profiles.  We are  also constructing lens
models  for the  other  clusters in  our  sample, and  will study  the
sub-structure and correlations  between cluster properties e.g. $L_X$,
$T_X$ and  mass (Smith  et al.\ 2002b).   In conjunction  with ongoing
\emph{XMM-Newton} observations,  this will be a  powerful dataset with
which to  quantify the distribution  of matter in massive  clusters at
$z\sim0.2$.

\section{The Surface Density of Extremely Red Objects at $K\ls22$}

We  also  use  these   clusters  as  ``gravitational  telescopes''  to
investigate properties  of distant  galaxies.  In particular  we probe
the formation  epoch of massive  elliptical galaxies.  Photometrically
selected evolved galaxies at $z>1$  are a promising tool with which to
tackle  this  issue.   However,  the  definition  of  this  population
(Extremely Red Objects -- EROs) in terms of a simple optical/NIR color
selection criterion  (e.g.\ ($R-K$)$\ge5.3$) produces  a heterogeneous
sample of  galaxies, containing both active, dusty  systems (e.g.\ Dey
et al.\  1999) and passive  elliptical galaxies (e.g.\ Soifer  et al.\
1999).   Attempts  to  study  EROs  in  a  systematic  manner  and  so
disentangle  this mixture have  also been  severely hampered  by their
intrinsic faintness ($R\gs23$, $K\gs18$).

Our survey overcomes this  problem, exploiting the magnifying power of
massive  galaxy  cluster lenses  to  enhance  the  sensitivity of  the
observations (S02a).  We combine our  optical HST data (\S2) with deep
ground-based  NIR data and  construct a  sample of  60 gravitationally
lensed EROs  with $(R-K)\ge5.3$ (26 which have  $(R-K)\ge6.0$) down to
$K\sim22$ in the source plane (Fig.~2).

\smallskip
\noindent\begin{center}
\begin{minipage}{50mm}
\psfig{file=fig2a.ps,width=50mm,angle=-90}
\end{minipage}
\hspace{10mm}
\begin{minipage}{50mm}
\psfig{file=fig2b.ps,width=50mm,angle=-90}
\end{minipage}
\end{center}
\smallskip
{\footnotesize{\noindent{\sc    Figure~2.}    --    Left:    Composite
$(R-K)$--$K$ color-magnitude  diagram for the ten  cluster fields used
in  our  search for  gravitationally  lensed  EROs (S02a).   Selection
criteria of  $(R-K)\ge5.3$ and  $(R-K)\ge6.0$ are indicated  by dashed
lines.   Right:  Number  counts   of  EROs  with  $(R-K)\ge5.3$  after
correcting  for   gravitational  lensing.   The   surface  density  at
$K\le21.6$  is $2.5\pm0.4$  arcmin$^{-2}$.   The slope  of the  number
counts changes at $K\sim19.5$ (see text).

}}
\smallskip

We  use detailed  lens models  (K96, S01a,  S02b) to  correct  our ERO
sample for gravitational amplification  (S02a), and show the corrected
surface density  of EROs in  Fig.~2.  These results agree  with recent
wide-field  surveys  ($K\ls19.5$), but  reveal  a  shallower slope  at
$K\gs19.5$.   Together with  color and  morphological  information, we
interpret  this as indicating  that EROs  are dominated  by elliptical
galaxies  at  $K\ls19.5$,  and  are  dominated  by  dusty  systems  at
$K\gs19.5$.

We also  compare our number  counts with Cole  et al.'s (2000  -- C00)
semi-analytic  model of  galaxy formation  (Fig.~2).  C00  use  a well
motivated prescription  of galaxy formation physics  to reproduce many
of  the  properties  of  the  local  galaxy  population  (e.g.\  local
$K$--band luminosity function).   Despite success at low-redshift, C00
under-predict  the observed ERO  counts by  approximately an  order of
magnitude  (Fig.~2).  This deficit  of EROs  implies that  C00's model
produces insufficient stars and/or dust in the early Universe.

Our future program  will test S02a's hypothesis that  the break in the
ERO number counts  is caused by a transition in the  nature of EROs at
$K\sim19.5$.

\section{ERO\,J164023 -- A Dusty Starburst-Seyfert Galaxy at $z=1.05$}

We  are also  undertaking a  program of  NIR spectroscopy  of  our ERO
sample.   Observations  with NIRSPEC  on  Keck-II  (S01b) reveal  that
ERO\,J164023  ($K=17.6$,  $(R-K)=5.9$)  is  a  dusty  disk  galaxy  at
$z=1.05$.   S01b  estimate  that  this galaxy  suffers  extinction  of
$A_V\sim5$, suggesting  that it  may be similar  to HR10 (Dey  et al.\
1999), one of only two other  dusty EROs for which NIR spectroscopy is
available.   However,  unlike  HR10,  this  galaxy  displays  evidence
(anomalous  H$\alpha$/[NII]  line  ratio)  of weak  nuclear  activity,
indicating  that this  may  be a  composite starburst-Seyfert  system.
Dust-reddened  EROs  therefore  appear  to include  composite  systems
(S01b) in addition to the starburst (Dey et al.\ 1999) and AGN (Pierre
et al.\ 2001) systems that were previously known.

\section{Summary}

We are  conducting a  survey of 12  X-ray luminous galaxy  clusters at
$z\sim0.2$ with \emph{HST} and  large ground-based telescopes.  One of
these clusters,  A\,383, contains  two radial arcs  which are  used to
constrain the inner  slope of the cluster density  profile (S01a).  We
find  that the cluster  scale dark  matter halo  appears to  possess a
central cusp.  However, the logarithmic slope of this cusp is not well
reproduced  by either  of the  rival theoretical  predictions  of dark
matter density  profiles.  S02a also  exploit the magnifying  power of
our cluster  lens sample to  construct a sample of  60 gravitationally
lensed EROs  as faint  as $K\sim22$.  Comparison  of our  observed ERO
number  counts  with predictions  from  C00's  semi-analytic model  of
galaxy formation reveals that  this models under-predicts the observed
ERO number counts by an  order of magnitude.  NIR spectroscopy reveals
that  one ERO in  our sample  is a  dusty starburst-Seyfert  galaxy at
$z=1.05$ (S01b).

\acknowledgements  I  acknowledge  all  of  my  collaborators  and  am
especially  grateful  to  Ian  Smail  and  Jean-Paul  Kneib.   I  also
acknowledge a Postgraduate Studentship from PPARC.

\end{document}